\documentstyle[12pt,aaspp4]{article}
\def\kms{km~s$^{-1}$~} 
\newcommand{\sso}{$^{-1}$}
\begin{document}
\title{Molecular Gas Kinematics in Barred Spiral Galaxies}
\author{Michael W. Regan,\altaffilmark{1,2}\\ 
\affil{Carnegie Institution of Washington,
Department of Terrestrial Magnetism,\\
5241 Broad Branch Road,
Washington, DC 20015}}
\author{Kartik Sheth\altaffilmark{3},
\&
Stuart N. Vogel\altaffilmark{4}}
\affil{Department of Astronomy,\\ University of Maryland,\\ 
College Park, MD 20742}
\altaffiltext{1}{Email-mregan@dtm.ciw.edu}
\altaffiltext{2}{Hubble Fellow}
\altaffiltext{3}{Email-kartik@astro.umd.edu}
\altaffiltext{4}{Email-vogel@astro.umd.edu}
\begin{abstract}
To quantify the effect that bar driven mass inflow can have on the evolution 
of a galaxy requires an understanding of the dynamics of the inflowing gas.
In this paper we study the kinematics of the dense molecular
gas in a set of seven barred spiral galaxies to determine 
which dynamical effects dominate.
The kinematics are derived from observations of the CO J=(1$-$0) line made
with the Berkeley-Illinois-Maryland Association (BIMA) millimeter array.
We compare the observed kinematics to those predicted by 
ideal gas hydrodynamic and ballistic cloud-based models
of gas flow in a barred potential.
The hydrodynamic model is in good qualitative agreement with both
the current observations of the dense gas and previous observations of the
kinematics of the ionized gas.
The observed kinematics indicate that
the gas abruptly changes direction upon entering the dust lanes to flow
directly down the dust lanes along the leading edge of the
bar until the dust lanes approach the nuclear ring.
Near the location where the dust lanes intersect the
nuclear ring, we see two velocity components:
a low velocity component, corresponding to gas on circular orbits, and
a higher velocity component, which can be attributed to the fraction of
gas flowing down the bar dust lane which sprays past the contact point
toward the other half of the bar.
The ballistic cloud-based model of the ISM 
is not consistent with the observed kinematics.
The kinematics in the dust lanes require large velocity gradients which
cannot be reproduced by an ISM composed of ballistic clouds with long
mean-free-paths.
Therefore, even the dense ISM responds to hydrodynamic forces.

\end{abstract}

\section{Introduction}
Mass inflow in spiral galaxies due to stellar bars has been 
used to explain a variety of phenomena in galaxy
evolution.
These phenomena include fueling starbursts (\markcite {HS94}{Heller \& Shlosman 1994})
and active galactic nuclei (\markcite {SFB89}{Shlosman, Frank, \& Begelman 1989}) and
even creating bulges in late-type spirals (\markcite {N96}{Norman, Sellwood,
\& Hasan 1996})
which may change their Hubble type to an earlier type and eventually
destroy the bar (\markcite {FB93}{Friedli \& Benz 1993};
\markcite {N96} {Norman et al. 1996}).
Even so,
direct observational evidence for mass inflow is
sparse.  The fundamental problems are that the gas flow is not
axisymmetric and only the line-of-sight velocity component is measured.  
As a result, to date evidence for mass inflow requires comparing 
observations to models to recover the missing velocity component.
This was done for the barred galaxy NGC 1530, where a good
match was found between an ideal gas hydrodynamic model
(Piner, Stone, \& Teuben 1995; hereafter PST95) 
and the H$\alpha$ velocity field (Regan, Vogel, \& Teuben 1997;
hereafter RVT97).
However, H$\alpha$ emission traces relatively
diffuse interstellar gas. 
It would be more convincing to show that
dense molecular gas, which probably constitutes the bulk of the gas
mass and which may reside in  ballistic clouds
that do not respond to hydrodynamic forces, 
also exhibits the kinematics predicted by hydrodynamic models.
By observing a sample of galaxies at a variety of
viewing angles it should be possible to infer, on average,
the missing velocity component and to test whether the hydrodynamic model
or ballistic cloud model 
can best reproduce the kinematics of the dense gas.

In this paper we present CO J=(1$-$0) observations of
seven barred spiral galaxies. 
The galaxies were selected based on their proximity, previous
CO detections, and strength of dust lanes in the Hubble atlas (Sandage 1961).
Global characteristics of these galaxies
are listed in Table \ref{galchar}.  Using this sample, we investigate the
kinematics of dense gas in bar dust lanes and in the crucial region
where the bar dust lanes approach the nuclear ring and compare the
observed gas kinematics to those predicted by hydrodynamic and
cloud-based models.

\section{Observations and Data Reduction}

We observed the  CO J=(1$-$0) line toward seven barred spiral galaxies
using the Berkeley-Illinois-Maryland-Association (BIMA) Array during the
period 1993 to 1997. 
The details of the observing are listed in Table \ref{coobs}.
For all the observations the instrumental gain and phase were calibrated 
from observations of a nearby quasar approximately every 30 minutes, 
and structure in the IF passband was calibrated with observations 
of a strong quasar.
The flux density scale was established by using observations of
either Mars or Uranus.
We estimate the uncertainty in the flux calibration to
be 30\%.

We formed naturally weighted maps of the CO data with 10 km s$^{-1}$ channels
and cleaned the map using the H\"{o}gbom algorithm.
We then performed an iterative phase-only 
multiple channel self-calibration
process as explained in Regan, Vogel, \& Teuben (1995).
The resulting phase corrections were not large ($<$ $\sim$30\arcdeg).
For each of the galaxies we determined the noise level, $\sigma$, from
emission free regions of the channel maps and
formed clipped red and blue total intensity maps
by summing all the
flux with $\mid S_\nu\mid >$ 2$\sigma$  on each side of the systemic
velocity (Figs 1-7a) within the velocity limits for each galaxy shown in Table \ref{coobs}.
We also formed a total intensity map from the channel maps for each galaxy
by summing all the flux over the imaged velocity range in Table \ref{coobs}
(Figs 1-7 b).
Fits versions of the data cubes for each galaxy have been placed in the Astronomy
Digitial Image Library.

NGC 3627 was reduced differently since it was observed as a mosaic of five pointings. 
In this case
we formed a linear dirty mosaic of the five pointings.
We then deconvolved the map using a maximum entropy method
(Sault, Staveley-Smith, \& Brouw 1996).
We used the resulting model for an iterative phase-only self-calibration.
We then primary beam corrected the map and formed the total intensity map.
This map is different from the other maps in that the noise level 
varies over the mapped region.

\section{Results}
\subsection{CO Morphology}

\subsubsection{NGC 1300}
NGC 1300 is a prototype strongly barred galaxy (Sandage 1961). 
Its HI velocity field has been studied by England (1989) and
Lindblad et al. (1997).
Optical images of the galaxy reveal straight dust lanes along the
leading edge of the bar and regions of active star formation at the
bar ends.
Optical surface photometry reveals an outer star forming ring that coincides
with the positions of large HI concentrations (Elmegreen et al. 1996). 

Due to the
low declination of the galaxy,
the resulting synthesized beam is elongated north-south and
the sensitivity is relatively low.
Figure \ref{conir1300}a shows that the CO emission is 
characterized by two peaks, a strong one west of the nucleus at location P and 
a broader weaker one east of the nucleus at location F.
There is a  weak extension of the CO extending out to the east 
from the east peak toward location D.

\subsubsection{NGC 1530}
NGC 1530 is one of the closest strongly barred northern hemisphere galaxies.
Molecular emission from NGC 1530 has been discussed by 
Regan et al. (1995), Downes et al. (1996),
Reynaud \& Downes (1997), and Reynaud \& Downes (1998).
In addition, the mass inflow rate into the central region has been estimated
to be 1 M$_{\sun}$ yr\sso\ from a comparison of H$\alpha$ Fabry-Perot 
observations and a hydrodynamic model (RVT97).

Figure \ref{conir1530}b shows that
the morphology can be characterized by two peaks of emission to the
northeast (location F) and southwest (location S) of the nucleus of the galaxy.
The emission at each peak is extended along the dust lanes.
Higher resolution maps (Regan et al. 1995; Reynaud \& Downes 1997)
reveal that there
may be an inner spiral or a broken ring of emission interior to the
two main emission peaks.

\subsubsection{NGC 2903}

NGC 2903 is a starburst galaxy that is classified as weakly barred in
the optical, although near infrared images (Regan \& 
Elmegreen 1997) reveal a strong bar.
Radio continuum, optical, and near infrared imaging 
of the nuclear region show 
a complex structure of ``hot spots'' of active star formation
(Wynn-Williams \& Becklin 1985; Jackson et al. 1991).
These hot spots are within the region of strong concentration
of molecular gas seen in single dish CO observations
(Jackson et al. 1991; 
Planesas, Colina, \& Perez-Olea 1997).

Figure \ref{conir2903}b reveals that
this galaxy does not show two peaks of emission at the inner 
ends of the dust lanes.
Instead, there is a central peak of emission whose spectrum 
contains a single broad velocity component centered on the
systemic velocity of the galaxy.
There is a broad extension of CO emission along the southern half of the bar
and a narrower extension along the northern half of the of the bar.
In addition, this galaxy shows a concentration of
emission at each end of the bar.

\subsubsection{NGC 3627}
NGC 3627 is member of the Leo triplet; NGC 3623 and the edge-on NGC 3628
make up the other two-thirds.  
The triplet shows strong evidence of interaction as indicated by an
HI tail 80 kpc east of NGC 3628 (Rots 1978; Haynes, Giovanelli, \&
Roberts 1979).
There is high 
star formation activity in both NGC 3627
and NGC 3628 based on radio continuum observations (Reuter et al. 1991)
and high IRAS flux densities (Rice et al. 1988; Young et al. 1989).
NGC 3627 is the most prominent member of the 
group with a pronounced dust pattern (Sandage 1961) 
indicating a strong spiral density pattern.
Measurements from the Kuiper Airborne Observatory 
at 50 and 100 $\micron$ by Smith et al. (1994) found the
far IR to H$\alpha$ ratio to be very high, implying large extinction in the
central region. 
NGC 3627 has been previously observed in CO$(1-0)$ and CO$(2-1)$ with the IRAM 30m (Reuter
et al. 1996) and with the 3-element BIMA interferometer in CO$(1-0)$ (Zhang, Wright, \&
Alexander 1993). Our new 9-element 
observations have a larger field of view and
greater sensitivity than the previous BIMA observations.

The CO map shown in Figure \ref{conir3627}a is
different than the others in this paper since it was
formed from a mosaic of five pointings.
This yielded a very large field of view allowing the global CO morphology
to be viewed.
As previously seen by Zhang et al. (1993),
the center of the galaxy shows a single emission peak on the nucleus of
the galaxy extending out into two offset ridges.
Both ridges connect to the large CO concentrations at the
ends of the bar.
The two emission peaks at the bar ends are quite strong and the CO can be
seen to connect to the spiral arms.
Here, as in NGC 2903, the CO emission in the central CO peak 
has only one velocity component centered on the systemic velocity of
the galaxy.

\subsubsection{NGC 4314}
NGC 4314 is an SBa galaxy with a
140\arcsec\ bar and two faint spiral arms tracing
130 degrees of arc out to a radius of 125\arcsec.
H$\alpha$ imaging of the galaxy shows it to have a nuclear ring of star formation
(Pogge 1989).
NGC 4314 has been observed in CO J=$(1-0)$
with the OVRO interferometer (Benedict, Smith, \& Kenney 1996).
HST observations reveal a ring of star forming knots
coincident with the H$\alpha$ ring (Benedict et al. 1993).

Figure \ref{conir4314}b shows that
there are two peaks of emission in NGC 4314.
They are not located at the
inner terminus of the dust lanes since
the dust lanes terminate to the northeast and southwest of the nucleus while
the two peaks are to the northwest and southeast.
Although it is not clear from our map, the higher resolution OVRO maps
show that the CO is distributed in a ring (Benedict et al. 1996).

\subsubsection{NGC 5135}
NGC 5135 has been called a mini-Seyfert 2 (Phillips et al. 1983; Thuan 1984).
It was detected in the
single dish CO survey of Heckman et al. (1989).
For our observations
the low declination of this galaxy yielded a very elliptical beam.
Figure \ref{conir5135}b shows that
the CO emission is only partially resolved by our beam and is centered on the
nucleus of the galaxy with a small extension visible to the southeast along
the leading edge of the bar.
The spectrum of the emission from the central region
shows a single velocity component centered on the systemic velocity of the 
galaxy similar to what is seen for the central regions of NGC 2903 and NGC
3627.

\subsubsection{NGC 5383}
NGC 5383 is a relatively nearby northern hemisphere barred spiral which has 
been used for 
many models of gas flow in barred potentials
(Sanders \& Tubbs 1980; Duval \& Athanassoula 1983).  
Optical studies have described the morphology of the galaxy in
detail (Burbidge, Burbidge, \& Prenderdast 1962; 
Duval \& Athanassoula 1983; Elmegreen \& 
Elmegreen 1985, Barton \& Thompson 1997). 
The galaxy is inclined at 50$^o$ and
has a prominent bar (110\arcsec) at a 
position angle of 135$^o$ (Sheth et al. 1998; Duval \& Athanassoula 1983), 
classic offset dust lanes and faint spiral arms.  
Kinematics of the galaxy have been studied using slit spectroscopy 
(Burbidge et al. 1962; Peterson et al. 1978; Duval \& Athanassoula 1983), HI interferometry 
(Sancisi, Allen, \& Sullivan  1979). 
Extensive hydrodynamic modeling of the galaxy has been done by Huntley (1978) and
Athanassoula (1992b). Orbital dynamics have been studied by 
Sanders \& Tubbs (1980), Tubbs (1982), and Athanassoula (1992a).  
This galaxy is the subject of an extensive multi-wavelength study in Sheth 
et al. (1998).

Figure \ref{conir5383}b shows that a large fraction of
the detected CO emission from NGC 5383 is in two peaks at the inner ends of the
dust lanes. 
In addition, it also shows an extended region 
of emission between the peaks of emission and a weaker
peak centered on the nucleus.
There is also faint emission extending from the two strong peaks along the dust lanes.

\subsubsection{CO Morphology Summary}

Three of the seven galaxies (NGC 1300, NGC 1530, and NGC 5383)
exhibit the ``twin peaks'' morphology described by Kenney et al. (1992).
This morphology is characterized by two emission maxima near the nucleus,
generally at the inner terminus of the straight offset bar dust lanes.
Three of the other four galaxies (NGC 2903, NGC 3627, and NGC 5135)
have essentially unresolved emission centered
on the nucleus.
The spectra of this emission show a single broad velocity component.
Only NGC 4314 shows a ring of molecular emission.

Another region in barred galaxies where we would expect to detect emission is from
the ends of the bars (Handa et al. 1990, Kenney \& Lord 1991, Downes et al. 1996).
The fact that we only detect gas at the ends of the bar in two of the seven
galaxies (NGC 2903 and NGC 3627) observed in this paper is not significant.
In all galaxies but NGC 3627 we observed at only a single position causing 
the flux from the bar ends to be severely attenuated by the primary beam
of the antenna.
Of the single pointing galaxies, only in NGC 2903 was the bar end both close 
enough to the center of the field and strong enough to detect.
Since single dish CO observations of NGC 1530 revealed concentrations of gas
at the ends of the bar (Downes et al. 1996), we know that such concentrations can
exist and not be detected in our maps.

\section{CO Kinematics}

RVT97 showed that the H$\alpha$ velocity field of
NGC 1530 is in
good agreement with
the global velocity field of a barred galaxy predicted by
hydrodynamic models (PST95).
Both the
 model and observations showed a strong shock along the entire
leading edge of the bar, isovelocity contours which become parallel
in the nuclear region, a 10\arcdeg\ twist in the inner kinematic
major axis, and kinks in the isovelocity contours in the spiral arms.
There are several potential problems with using H$\alpha$ as a
tracer of the
ISM kinematics:
H$\alpha$ emission arises from the diffuse gas in the ISM,
can be biased by regions of active star formation,
and is subject to extinction by dust.
We would like a tracer of the gas that does not suffer
these biases and one that 
traces the dense molecular gas that probably constitutes the bulk of the gas
mass.
Carbon monoxide is just such a tracer, allowing 
it to provide independent confirmation of the model.
Because CO emission only arises from regions where the ISM is molecular,
we can determine what we would expect to observe
for the CO kinematics
by looking at the kinematics of the dense gas
in the model.

We have chosen to focus on comparing the kinematics to the models
 rather than the
morphology because
 the models that we are using do not attempt to take into
consideration any consumption of gas by star formation.
Since the removal of gas over time can be quite large, we would
not expect the
morphology of the models and observations to match.
On the other hand, the kinematics should not be changed by the removal of
gas by star formation and so are a better test of the models.

\subsection{Hydrodynamic Model Predictions}

Figure \ref{hydroden} is a plot of the gas surface density 
from the high resolution hydrodynamic model (run 4)
of PST95 near the end of the simulation after it
has reached a nearly steady state configuration.
Three different regions are labeled in the figure:
the leading dust lane, the contact
point where the dust lane meets the nuclear ring, and a
spray region downstream of
the contact point and the nuclear ring where the gas flow is
divergent.

Figure \ref{hydrovel} shows the model gas velocity field in the plane of the 
galaxy in a reference frame rotating with the bar.
From the figure we can see that
the high density gas in the dust lane is moving directly down the
dust lane. At the contact point
the gas in the nuclear ring is moving on nearly circular
orbits, while the gas just outside of the ring is moving faster than the
gas in the ring.
Finally, in the spray region the gas that
is spraying back out into the bar is moving
faster than the gas in the nuclear ring.
Here gas is in a divergent flow as it leaves the contact point.

Since we do not observe the velocity in the plane of the galaxy but only
the line-of-sight velocity, the observed velocity of the gas will
be affected by the orientation of the galaxy on the sky.
Along the galaxy major axis we only observe velocity that is tangential
in the plane of the galaxy.
While on the galaxy minor axis, we only observe motion that is radial in 
the plane of the galaxy.
Moreover, the sign of the
projected galactic radial motion depends on whether the observed
location is on the near or the far side of the galaxy.
For example, on
the far side, radial inflow is observed blue-shifted relative to the
systemic velocity, while on the near side it is red-shifted.
Therefore, interpretation of the kinematics requires knowledge of
both the galaxy position
and which side of the galaxy is the near side.
For all of the galaxies in this paper, we use the usual assumption that the
spiral arms are trailing to allow us to determine the near side of the
galaxy and the convention that negative radial velocities 
in the plane of the galaxy are inward.

Consider first gas in the bar dust lane located far from the
nuclear ring.  Figure \ref{hydrovel} shows that in the bar frame such gas has very
little tangential motion but a large radial inward velocity.  The
additional tangential motion resulting from the bar rotation
(this ranges from 20 km s\sso\ kpc\sso\ to 60 km s\sso\ kpc\sso (
Teuben et al. 1986; 
Buta 1986; 
Jorsater \& van Moorsel 1995;
England, Gottesman, \& Hunter 1990;
Merrifield \& Kuijken 1995;
Ryder, Buta, \& Toledo 1996;
Buta \& Purcell 1998))
can generally be estimated. 
However, we note
that if the bar dust lane lies close to the minor axis then there is
no Doppler contribution from bar rotation, and motion along the dust
lane is directly observed.  On the other hand, if the observed
location is close to the major axis, then the observed velocity
depends little on the flow speed along the dust lane, but rather on the
pattern speed of the bar.

Consider second gas in the bar dust lane located near the nuclear ring
(i.e. near the contact point).  Figure \ref{hydrovel} shows that 
in this
case gas is flowing nearly tangentially.  Due to beam-smearing, emission
observed at this location comes not only from this dust lane gas, but
also from gas rotating in the nuclear ring.  The latter gas is also
moving tangentially but at a lower speed than the dust lane gas.
Therefore, near the contact point we expect spectra to show two
components, corresponding to the dust lane gas and the nuclear ring
gas.  The nuclear ring gas has a lower tangential speed and will
therefore be the component closer to the systemic velocity of the
galaxy.  In some cases, the velocity separation between the two
components will be too small to resolve the two components.

Finally, consider gas in the spray region (i.e. downstream from the
contact point).  This gas has both an outward radial velocity and a
tangential velocity that is greater than or equal to the
gas in the nuclear ring.
Again given the finite resolution of the CO observations we would expect that
the spraying gas spectra will be superimposed upon the spectra of the
denser gas in the nuclear ring. 
The relative velocities of these two components will depend upon the projections
in the specific galaxy.

\subsection{Cloud-based Model Predictions}

An alternative to the hydrodynamic approach to modeling the ISM 
is to assume that the ISM is
composed of an ensemble of colliding gas clouds that move on ballistic orbits
(Schwarz 1984, Roberts \& Hausman 1984, Combes \& Gerin 1985, 
Palous, Jungwiert, \& Kopecky  1993,
Byrd et al. 1994).
In this approach the ISM does not experience any pressure forces and thus 
there are no large-scale shocks.
Due to the collisions in the cloud ensemble, the cloud orbits will
lead the bar within the bar radius (Combes 1996), resulting in orbits such as those
shown in Figure \ref{gasorbits}.

From these gas orbits we can make predictions about what the 
observed kinematics should be if the ISM is well approximated
by an ensemble of gas clouds.
First, consider the gas clouds in the 
bar dust lanes far from the nuclear
ring. 
Figure \ref{gasorbits} shows
that the dust lane is formed from the crowding
and superposition of several orbits with small differences in
total energies.
All of these velocities will have the same direction since the
various orbits are nearly parallel in the dust lanes.
The tangent vector to the orbits in the dust lane shows that
the clouds in the dust lanes are not moving as directly inward
as the gas in the dust lanes shown in Figure \ref{hydrovel}.
This 30$-$40\arcdeg\ difference in the direction of the gas flow
would be difficult to detect unambiguously.
The biggest difference in the predicted kinematics in the dust lane
comes in the gas that is upstream of the dust lane.
In the cloud-based model there is not a shock or a large change in
gas velocity in the dust lane. 
The gas upstream of the dust lane has almost the same speed and
direction of flow as the gas in the dust lane.
It is flowing inwards with low but non-zero tangential velocity.
This is in contrast to the gas upstream of the dust lane in the
hydrodynamic model where the gas has a much different direction of
flow from the gas in the dust lanes.

While the contact point is not well defined in the cloud-based model,
the expected kinematics at the observed contact point 
are similar to those in the hydrodynamic model. 
We would expect to see slow moving gas in the nuclear ring superimposed
on the faster moving clouds that are moving past the ring.

Similarly, in the spray region Figure \ref{gasorbits} shows that the gas
orbits are diverging and heading around to contact the dust lane on the
other side of the bar. 
Although the gas orbits are not diverging from a single point as
the gas streamlines do in the hydrodynamic model, the differences are
very small.
Therefore, the expected kinematics are the same for the two models of
the ISM in this region.

\subsection{Observed Kinematics}

To investigate the gas kinematics we have chosen to focus on the
gas as it flows down the dust lane. 
The dust lanes and nuclear region are where the ISM density is
high enough to detect molecular gas.
The dust lanes are also the critical region 
for the predictions of the hydrodynamic
model since the dust lanes are where 
the cloud-based and hydrodynamic models
predict different kinematics.
For each dust lane of each galaxy we extracted a series of spectra along the leading
edge of the bar extending to the contact point and continuing out into the spray region.
These locations may not exactly match the locations of the observed dust lanes but 
allow us to use a consistent set of locations for all the galaxies.
The differences between where the spectra are extracted and the actual locations of the
dust lanes are small relative to the sizes of the synthesized beams.
Each spectrum is separated by approximately the size of the synthesized
beam.
The galaxy position angles used for this analysis are shown in Table \ref{galposang}.
The locations of each spectrum,
the galaxy axes, approaching and receding sides,
and near and far sides are shown in Figures 1-7b.

\subsubsection{NGC 1300}

These spectra have a low signal to noise ratio which when combined with the
large elliptical synthesized beam leads to spectra that are
difficult to interpret.
Even so, at several locations we detect kinematic features that 
are significant.
At location P we see gas that is close to the major axis of the
galaxy and in a dust lane.
The spectrum at this location shows 
two velocity components.
One is about 70 km s\sso\ wide
centered on the systemic velocity
and the other is red-shifted relative to the systemic velocity of the
galaxy.
The component centered on the systemic
velocity must have very little tangential velocity;
this is consistent with the inflowing gas in the dust lanes.
The other component seen at this location 
must have significant tangential velocity;
this is consistent with this being gas in 
near-circular rotation at the nuclear ring.
Presumably the two components are spatially distinct but
are not resolved by our beam.

The spectrum at location H, which is near the 
spray region of the eastern dust lane and the nuclear ring, shows 
weak broad blue-shifted emission with velocities
of 100 km s\sso\ relative to the systemic velocity.
Since this location is approximately on the near side minor axis, this gas
may be flowing radially outward in the plane of the galaxy
or it may be that emission here is beam-smeared from the other side of
the galaxy.
There is also some gas red-shifted as much as 50 km s\sso\ relative to 
systemic velocity but this is clearly due to emission beam-smeared from the
the strong emission seen at location P. 
The relatively large synthesized beam means that an unambiguous interpretation
of the spectrum at this location is not possible.

\subsubsection{NGC 1530}

In the eastern dust lane of NGC 1530 we first detect faint emission at location
A.
There is weak emission near systemic and at 2580 km s\sso\ which is about
120 km s\sso\ red-ward of systemic.
At this location between the receding major axis and the far side minor
axis, only if the gas has a significant radial inward velocity will it 
have an observed velocity near systemic.

Moving in
along the eastern dust lane of NGC 1530 we detect emission
at location C.
Here the major component is slightly offset to the blue from the systemic
velocity. 
This position in NGC 1530 is about 30\arcdeg\ from the
far side minor axis on the receding side of the galaxy.
Therefore, gas in circular orbits will be red-shifted relative to the
systemic velocity.
For gas to be at the observed blue-shifted velocity it would have to have some 
combination of slower than circular tangential velocity and inward radial
motion.
Since this is in the dust lane, that is exactly the effect we expect to
see.
In addition,
a weak narrow
detection of gas at near circular velocities is seen at 2570 km s\sso.
The observed spread in velocities at this location (150 km s\sso) would
be hard to explain if the ISM were composed of ballistic clouds.

Location E is very close to the far side minor axis of the galaxy.
Here we see very broad emission from 2320 km s\sso\ to 2520 km s\sso.
Since this location is on the far side of the galaxy 
inflowing gas will be
blue-shifted and outflowing gas will be red-shifted.
The observations show that the majority of the gas is inflowing.
The inflowing gas is consistent with our expectation of what gas should be
doing in the dust lane.
The small amount of red-shifted emission that we see is
due to beam smearing of gas from the receding side of the galaxy

Location F is at the contact point of the dust lane and the nuclear ring.
This location is also close to the major axis.
The spectrum shows that all of the gas is blue-shifted with
the strongest component centered on 2350 km s\sso.
A second possible component, at a more blue-shifted velocity
than the majority of the gas, is seen at a velocity of around 2280 km s\sso.
This second component is what would be expected for the higher velocity
gas that is streaming past the nuclear ring.

The existence of this higher velocity component is confirmed by the spectrum
at location H.
This spectrum is on the near side of the galaxy close to the major
axis.
There is clearly gas at velocities from 2260 km s\sso\ to 2320 km s\sso.
This is gas in the spray where we expect the gas to have a high
tangential velocity and be moving outward from the center of the galaxy.
Given the projection of NGC 1530, these two effects will add constructively
so we see this as a blue-shifted velocity component relative to the gas
on circular orbits in the nuclear ring.

In the western dust lane of NGC 1530 the first significant emission that
is detected is at location M.
Here we detect broad emission extending from systemic to 100 km s\sso\ 
blue-ward.
Emission over such a large range in velocity is not consistent with the
ISM in the dust lanes being composed of primarily dense clouds.

Moving inward toward the nucleus significant emission is seen at location R.
This location is the mirror image of location E in that it is on the near side
minor axis.
The spectrum looks like the reflection of the spectrum at location E as well.
The majority of the emission is red-shifted relative to the systemic velocity
with emission detected out to 2640 km s\sso\, 
approximately 170 km s\sso\ from systemic.
This red-shifted gas is flowing inward in the plane of the galaxy.
The small amount of emission that is seen blue-ward of systemic velocity
is probably due to beam smearing of gas from the ring region.

Location S is at the contact point of the dust lane and the nuclear ring.
This location is about midway between the near side minor axis and the
red major axis.
We expect gas to be moving completely tangentially here, and the large amount
of gas that is red-shifted relative to systemic is consistent with that.
We do not detect the expected higher velocity component here.
The gas that is seen around systemic velocity is unexpected. 
If this were emission from gas at location S then the gas would be flowing
outward.

At location T we are slightly downstream of the contact point and on the
major axis of the galaxy.
Here we see that the red wing of the 
velocity profile shows a component that is rotating
much faster.
This is consistent with the fast moving gas that is moving
past the nuclear ring.

Finally, at location U, we are well past the contact point and in the spray 
region.
Here the gas is slightly on the far side of the major axis.
The high velocity component seen at location T is not seen here although we
do see emission at velocities up to 2610 km s\sso.

\subsubsection{NGC 2903}

Although the emission in NGC 2903 is strong, the central emission is 
very close to
the nucleus of the galaxy and the synthesized beam does not resolve
it.
This leads to large amounts of beam smearing resulting in large velocity 
widths.

At location B on the the north side of the galaxy we do detect gas at the
end of the bar.
Since in NGC 2903 the bar major axis is closely aligned with the 
major axis of the galaxy, along the bar we are observing mainly
motion that is tangential in the plane of the galaxy.
The observed velocity of the 
gas at location B is only 60 km s\sso\ from systemic. 
This observed velocity is small compared to the observed HI line width of 
380 km s\sso 
(de Vaucouleurs et al. 1991).
Since at this location we are observing gas 2.5 kpc from the nucleus we would expect that
the circular velocity would be a larger fraction of the total line width.
Location B is on the near side of the galaxy so inflowing gas will be red-shifted.
Therefore, this component must have some combination of lower than circular
tangential velocity and inward radial velocity.
Further inward
at location F in the northern dust lane
there is gas at velocities close to the systemic velocity.
Since F is close to the near-side minor axis 
this is what would be expected for gas 
on circular rotation.
At F we also see a component that is blue-shifted implying
either fast tangential motion or outward flowing gas.
This emission is probably due to 
beam smearing from either the nuclear region or from the
nearby major axis.

Since we do not see the standard twin peak morphology in NGC 2903,
the contact point and the spray region are not easy to locate.
Location G, which is past the contact point of the 
northern dust lane, is between the near side 
minor axis of the galaxy and the receding major axis. 
We can
see that there is gas both blue-ward and red-ward of systemic.
The relatively large size of the synthesized beam compared to the 
galactic radius at this location means that the spectrum 
contains emission from a variety of projections making it impossible
to interpret.

In the southern dust lane we also detect gas near the
bar end in at location M.
This gas, like the gas at location B, 
is rotating too slowly to be in a circular orbit.
In the dust lanes at location P we detect
emission extending over a broad range of velocity (520 km
s\sso\ to 700 km s\sso) with a peak near systemic.
Emission over such a large range of velocity near the bar dust lanes
is not consistent with the expectations of an ISM composed of ballistic 
clouds.
Since location P is on the far side of the receding major axis,
circular motion should be red-shifted.
Emission could approach systemic by having a combination of
low tangential velocity and inward radial velocity.

This same effect (low tangential velocity and inward radial motion)
is seen at Q where the emission is blue-shifted from systemic.
Since Q is about 60\arcdeg\ from the minor axis 
and NGC 2903 has an inclination of
60\arcdeg, the projected
radial component is only 43\% of the velocity in the plane of the galaxy.
Thus, for gas to be seen 40 km s\sso\ blue-ward of systemic requires the gas
to be flowing inward with a velocity of at least 90 km s\sso.
Note that this is a lower limit because any tangential component of velocity
will red-shift the emission requiring an even greater inflow speed to yield
the same observed velocity.

At location R there is a broad emission profile extending to both sides
of the systemic velocity.
As with location Q, in order 
for there to be emission blue-ward of systemic requires
the gas to be flowing radially inward.
Gas here is detected 60 km s\sso\ blue-ward of systemic requiring a minimum
inflow speed of 130 km s\sso.

\subsubsection{NGC 3627}

Along the southern dust lane of NGC 3627 we see emission all the way from 
the bar end to the nuclear region.
Location M is near the end of the bar and the spectrum at this location
shows a rotation velocity centered around 910 km s\sso.
Location P is on the major axis of the galaxy and in the dust lane.
The spectrum at this location shows an emission peak around 790 km s\sso\ with
a large red wing to the emission.
Given the higher rotational velocity seen at location M, the slower rotating
gas at location P in the dust lane must
have a significantly lower tangential velocity consistent with gas flowing directly
down the dust lane.

The fact that there is also gas at location P with significant tangential velocities
implies that we are detecting gas before it enters the dust lane while it still 
has significant tangential velocities.
This large change in velocities at the dust lane is not consistent with the expectations
of the cloud-based model.

At location Q 
we are looking just off the major axis on the far side of the galaxy.
Here we see a spectrum similar to the one at location P.
Therefore, the bulk of the gas is not rotating fast enough to be in
circular orbits.

In the northern dust lane we detect emission near the bar end at location A.
Similar to location M in the southern dust lane, except that the emission is
offset to the blue, 
here we can see that the rotational
velocity is centered around 560 km s\sso.
Further down the dust lane at locations E and F we can see that at these
locations, which 
are very near the major axis, the gas is rotating much slower than at A.
Both locations have emission very near to systemic.
Given the minimum tangential velocity provided by the rotation of the bar,
the bulk of the gas at location F must have a very small tangential component
of velocity in the rotating reference frame of the bar.
Again the large observed line width at location F is not consistent with
the expectations of a ballistic cloud-based ISM. 

The nuclear region of NGC 3627 is not resolved by our synthesized beam.
Therefore, it is not possible to determine where the contact point of 
the dust lane and the nuclear ring is.
We can see that spectra the include the minor axis (locations H and S) reveal
emission from a large range in velocity. 
This arises from the large beam relative to the galactic radius. 
This causes the spectrum to include emission from many different projections.
much is due to beam smearing. 
Since we expect the gas in the ring and the gas spraying past the ring to be
moving purely tangentially here, observations of the galactic radial velocity
should not reveal two components.
The galaxy is not favorably oriented for the
spectra in the region where there would be a spray (locations U and J)
to show the spray component.
This is because the spray should have high tangential velocity and outward
radial velocity.
The orientation of NGC 3627 is such that at locations U and J these two effects
have opposite Doppler effects and thus the spraying gas is hard to
differentiate from gas on circular orbits.

\subsubsection{NGC 4314}
Our map of NGC 4314 does not detect any gas in the dust lanes.
At the location where the dust lane joins the nuclear ring 
in the north, location D, the spectrum shows a single broad velocity
component with velocities on both sides of systemic.
Since location D is between the far-side minor axis and the receding
side major axis, circular rotating gas will be red-shifted and inflowing
gas will be blue-shifted.
Therefore, the only way to get gas with velocities blue-ward of systemic is
for the gas to be flowing inward.
Similarly, at the other dust lane terminus, location Q, where we are
observing gas very near the approaching major axis,
we see a red wing to the emission profile indicative of gas
with a combination of slow tangential velocities and inward radial
motion.
In a more sensitive OVRO map of NGC 4314, Benedict et al. (1996) also show 
that at these locations gas is streaming
radially inward.
In the expected spray region (locations T and F) we do not detect any
significant emission.

\subsubsection{NGC 5135}

Interpretation of the kinematics in NGC 5135 is difficult
due to the elliptical beam that blends the kinematics.
Even so, there are several features that can be unambiguously detected.

At location D in the dust lane we are looking at gas that is between the
near-side minor axis and the approaching major axis.
Since this is on the near side of the galaxy, gas that is moving inward radially
will have emission red-shifted toward systemic.
Because this location is between the major and minor axis we are seeing
the blending of radial inflow and tangential rotation.
The gas in the dust lane is flowing inward with very little tangential velocity.
The radial inflow and low tangential velocity
combine to make the gas have more red-shifted emission 
than gas on circular orbits would have.
In fact, that is what we see in the spectrum. There is one spectral feature
at a velocity of approximately 4020 km s\sso\ and another stronger feature
at a velocity of 4060 km s\sso.
This second feature even has a red wing that extends red-ward of systemic.
Gas can be observed with these velocities 
only if it is moving radially inward.

Location E is near the contact point of the southern dust lane and 
again is between the near-side minor axis and the approaching major axis.
Here again we are seeing the blending of tangential and radial motions.
The velocity profile is asymmetrical with gas in the blue wing having high
relative tangential motions.
There is also gas red-ward of systemic which could be indicative of inflowing 
gas.
Unfortunately, the large elliptical beam means that the spectrum at location
E is from a very large area including the receding side of the galaxy.
Therefore, it is not possible to definitively state that there is inflowing gas
at this location.

In the northern dust lane we detect gas at location P.
Here we are looking near the receding major axis 
on the far side of the galaxy.
We see gas here with a velocity that is blue by as much as 
50 km s\sso\ relative to systemic here.
Only by flowing radially inward can gas be seen blue-ward of systemic at this
location.
Therefore, the gas here must have a low tangential velocity and be streaming
radially inward.

\subsubsection{NGC 5383}

In NGC 5383 we do not detect any CO in the dust lanes except near 
where  the dust lanes join the ring.
At Location B on the western side of the galaxy we detect the
first significant emission.
At this location inflowing gas will be blue-shifted due to its
inward radial motion and red-shifted toward systemic due to its low
tangential motion.
Therefore, these two Doppler effects compete against each other making it
hard to determine whether we are seeing inflow.

Location D, at the contact point of the dust lane and the nuclear ring, is almost
on the major axis of the galaxy;
therefore,
we are observing mostly tangential motion in the plane of the galaxy.
From the hydrodynamic 
model we would expect to see the gas in the ring rotating at near
circular velocity. 
This component is seen to peak at a velocity of 2180 km s\sso.
A second faster moving component is also seen with its velocity 
centered at around 2110 km s\sso. 
This is, presumably, the gas that is streaming past the ring.

As we follow the dust lane out into the spray region, we
can see this higher velocity component remaining.
At locations F and G, well into the spray region, we detect gas with a
velocity of around 2110 km s\sso. 
At this location we are on the near side of the galaxy so the expected
radial outflow will be blue-shifted relative to the systemic velocity.
The expected higher tangential component will also be blue-shifted so
the two Doppler 
effects combine here allowing a clear detection of the spraying gas.
The fact that the faster moving component is stronger relative to the 
slower moving component is evidence that we are not just seeing emission
blending out to this location from the stronger emission at location D.
This is because
at location D the slower moving component is much stronger than the
faster moving component.

On the eastern side of the galaxy we detect gas in the dust lane as it
nears the ring.
Here too, the projections of the
combination of low tangential motion and radially inflowing
gas interfere, making it impossible to 
determine if the gas motion in the dust lane
is directly down the dust lane.

At the contact point of the dust lane and the ring, location P, we do
see a red wing to the emission.
Since on this side the contact point is also
 close to the major axis, this gas in the 
red wing of the profile is gas with a higher tangential velocity.
Again we are seeing the gas that is moving past the ring.

As we move out into the spray region we also detect a higher velocity
component to the emission.
On this side of the galaxy the combination of radial outflow and high
tangential velocity combine to red-shift the gas as 
we see at locations Q, R, and weakly at S.
This is the gas that is spraying around to contact the western dust lane.

\section{Discussion}
Overall, the hydrodynamic model is in better agreement with 
the CO kinematics than the ballistic cloud model.
In the dust lanes we detect gas with low tangential velocity in NGC 1300,
NGC 1530, and NGC 3627. We detect gas that has some combination of
radial inflow and low tangential velocity in NGC 2903, NGC 4314 and NGC 5135.
In NGC 5383 the projection of the galaxy precluded us from 
distinguishing streaming motions from circular motions in the dust lanes.
Therefore, the prediction of the hydrodynamic model
that the gas in the dust lanes is flowing directly
down the dust lanes is consistent with all of our observations.

Recently more sensitive observations of NGC 1530 made by combining observations
from the IRAM interferometer with observations from the IRAM 30m confirm
our interpretations (Reynaud \& Downes 1998).
These observations show that CO is moving 50 to 100 km s\sso\
radially outward upstream of the dust lanes and 70 to 150 km s\sso\ radially
inward in the dust lanes.

The expectations of an ISM dominated by ballistic clouds are not consistent
with either
our observations of NGC 1530, NGC 2903, and NGC 3627 
or previous ones of NGC 1530 (RVT97; Reynaud \& Downes 1998).
The sharp decreases seen in the tangential velocity at the dust lanes
cannot be explained
by collisions of clouds with long mean-free-paths.
Instead the ISM in this region seems to be well represented by an ideal
gas approximation.
This does not preclude the molecular gas from being in clouds.
It only implies that the clouds must respond to hydrodynamic forces
since the velocities of the dense gas change quickly upon entry to the dust lanes.

At the contact point we detect a higher velocity component in
NGC 1530 and NGC 5383. 
These are the only two galaxies in the sample where the contact point is
near the major axis of the galaxy.
Since the expected higher velocity component is moving completely tangentially
in the plane of the galaxy, only when we observe the contact point near the
major axis of the galaxy will we be able to detect the higher velocity component.
Therefore, given the projections, the expectations of the hydrodynamic model are consistent with
our observations.

The gas at the expected velocities of the spray region is found in 
NGC 1530 and NGC 5383.
Since gas in the spray region is flowing outward and has a higher than circular
tangential velocity, only at certain orientations will it be distinguishable
from the much stronger emission arising from the nuclear region.
Only if we observe the spray region in the receding far side or approaching
near side quadrants of a galaxy
will the projection cause the outward radial and faster tangential
Doppler effects to have the same sign.
Otherwise, the two Doppler
effects will have opposite signs and it will not be possible
to distinguish spray emission from ring emission.
In every galaxy where the projections are favorable
we do detect the spray.

The good agreement seen here between the hydrodynamic model 
and the observations has
important implications for modeling of gas flow in barred galaxies.
It was possible that the good fit to the hydrodynamic
model obtained from the
H$\alpha$ observations of NGC 1530 (RVT97)
could have been due to the fact that
the H$\alpha$ was tracing the diffuse ISM while the bulk of the ISM
remained in clouds that were not correctly modeled by an
ideal gas approximation.
Our current CO observations show that the ideal gas model also matches 
well the
kinematics of the dense gas
and that the ballistic cloud approximation of the ISM is not correct
within the region of the bar.

Higher resolution and more sensitive observations are needed to
confirm these suggestions.
The projections of the velocities change quickly in the nuclear regions
and observations at higher resolution would decrease the
beam smearing problem.
For example, the synthesized beam for our observations of NGC 5135 has
linear dimensions of 3 x 1 kpc. 
This creates a large ambiguity in where the actual location of the emission.

\section{Conclusions}

We have shown that a hydrodynamic model of
gas flow in barred galaxies is consistent with
the kinematics of the molecular gas in this set of
seven galaxies.
The gas 
quickly changes velocity upon entering
the dust lanes along the leading edge of the bar to flow
directly down the dust lane.
At the contact point of the dust lane and the nuclear ring,
when the projections are favorable, we see that
there are two velocity components. 
One is the slower rotating component that is in near circular orbits around
the nucleus. 
The other is the faster flowing gas that traveled down the dust lane and is now
flowing back into the bar region.
We have clearly detected this gas spraying back into the bar region in
those galaxies with orientations favorable for detection.

We have also shown that  a dense ISM consisting of ballistic clouds
is not consistent with the observed kinematics in the dust lanes.
The observed velocity line widths in the dust lanes are instead
consistent with the
dense ISM undergoing a shock similar to the one the diffuse ISM is experiencing
in NGC 1530 (RVT97).
These observations do not preclude the dense ISM from consisting of clouds.
They only require that the clouds respond to the hydrodynamic forces that
operate on an ideal gas.

\acknowledgements
The authors would like to thank John Lugten for the reduction of 
the NGC 5135 data.
They would also like to acknowledge helpful discussions with
Peter Teuben and Jim Stone.
Support for this work was provided by NASA through Hubble Fellowship
grant HF-01100.01 awarded by the Space Telescope Science Institute, which
is operated by the Association of Universities for Research in Astronomy,
Inc., for NASA under contract 5-26555.
We would like to thank the referee for comments that improved the paper.

\clearpage
\begin{figure}
\caption{NGC 1300 
a) CO total intensity (contours) overlaid on a near 
infrared image K-band (gray-scale) from Regan and Elmegreen (1997).
The grey contours are the intensity from the receding side of the
galaxy and the white contours are from the approaching side of the
galaxy.
The dotted line circle shows the  half-power radius of the primary beam, and
the synthesized beam is shown in the lower right.
b) CO total intensity for the central region of NGC 1300.
The bottom left of each letter marks
the locations at which the spectra presented in c) were
derived.
The region displayed is outlined with a rectangle in part a). 
The contours are at 2, 4, 6, and 8 times the noise in the map which is
4.3 Jy km s\sso.
The solid line is the major axis of the galaxy with the thicker half being
the receding side and the thin line the approaching side.
The dashed line represents the minor axis with the thicker line being the
far side.
c) Spectra of CO emission from NGC 1300 derived from the bottom left corner
of the letters plotted in b).
The systemic velocity is represented by the dotted line.
The resolution is 10 km\sso, and the noise in each channel is 88 mJy.}
\label{conir1300}
\end{figure}
\begin{figure}
\caption{
Same as Figure 1 except where noted.
NGC 1530 
a) CO total intensity (contours) overlaid on a near 
infrared image K-band (gray-scale) from Regan et al. (1995).
b) CO total intensity for the central region of NGC 1530 with the
locations of the spectra marked.
The contours are at 2, 4, 6, and 8 times the noise in the map which is
4.3 Jy km s\sso.
c) Spectra of CO emission from NGC 1530 at the locations marked in b).
The noise in each channel is 49 mJy.}
\label{conir1530}
\end{figure}
\begin{figure}
\caption{
Same as Figure 1 except where noted.
NGC 2903
a) CO total intensity (contours) overlaid on a near 
infrared image K-band (gray-scale) from Regan and Elmegreen (1997).
b) CO total intensity for the central region of NGC 2903 with the
locations of the spectra marked.
The contours are at 2, 4, 6, and 8 times the noise in the map which is
2.5 Jy km s\sso.
c) Spectra of CO emission from NGC 2903 at the locations marked in b).
The noise in each channel is 48 mJy.}
\label{conir2903}
\end{figure}
\begin{figure}
\caption{Same as Figure 1 except where noted.
NGC 3627 
a) CO total intensity (contours) overlaid on a near 
infrared image K-band (gray-scale) from Regan and Elmegreen (1997).
The dotted line circles show the  half-power points of the primary beams
of each of the pointings for the mosaic.
b) CO total intensity for the central region of NGC 3627 with the
locations of the spectra marked.
The contours are at 2, 4, 6, 10, 15, 20, 25, 30, 35, 40, 45, 50, 55,
60, 65, 70 times the noise in the map, 3.2 Jy km s\sso.
c) Spectra of CO emission from NGC 3627 at the locations marked in b).
The noise in each channel is 50 mJy.}
\label{conir3627}
\end{figure}
\begin{figure}
\caption{Same as Figure 1 except where noted.
NGC 4314
a) CO total intensity (contours) overlaid on a near 
infrared image K-band (gray-scale) from Regan and Elmegreen (1997).
b) CO total intensity for the central region of NGC 4314 with the
locations of the spectra marked.
The contours are at 2, 4, 6, and 8 times the noise in the map which is
1.8 Jy km s\sso.
c) Spectra of CO emission from NGC 4314 at the locations marked in b).
The noise in each channel is 38 mJy.}
\label{conir4314}
\end{figure}
\begin{figure}
\caption{Same as Figure 1 except where noted.
NGC 5135
a) CO total intensity (contours) overlaid on a near 
infrared image K-band (gray-scale) from Mulchaey, Regan, \& Kundu (1997).
b) CO total intensity for the central region of NGC 5135 with the
locations of the spectra marked.
The contours are at 2, 4, 6, and 8 times the noise in the map which is
3.8 Jy km s\sso.
c) Spectra of CO emission from NGC 5135 at the locations marked in b).
The noise in each channel is 84 mJy.}
\label{conir5135}
\end{figure}
\begin{figure}
\caption{Same as Figure 1 except where noted.
NGC 5383 
a) CO total intensity (contours) overlaid on a near 
infrared image K-band (gray-scale) from Regan and Elmegreen (1997).
b) CO total intensity for the central region of NGC 5383 with the
locations of the spectra marked.
The contours are at 2, 4, 6, and 8 times the noise in the map which is
3.3 Jy km s\sso.
c) Spectra of CO emission from NGC 5383 at the locations marked in b).
The noise in each channel is 54 mJy.}
\label{conir5383}
\end{figure}

\begin{figure}
\caption{Gas density from hydrodynamic model.
The three regions of interest, the dust lane, the contact point of the dust lane
and the nuclear ring, and the spray region are marked.}
\label{hydroden}
\end{figure}

\begin{figure}
\caption{Gas velocity from the hydrodynamic model in the rotating reference
frame of the bar.
The background gray-scale is the gas surface density.
The arrow represents the direction and magnitude of the gas
velocity.
Notice how the gas makes an abrupt change in direction when it
enters and flows down the dust lane.
At the contact point all of the gas is flowing tangentially and it is
clear that the gas in the ring (white arrow) is moving slower than the gas just outside of
the ring.
In the spray region the gas flow is diverging from the contact point.}
\label{hydrovel}
\end{figure}

\begin{figure}
\caption{Orbits of gas clouds in a barred potential under the assumption that
the ISM is composed of ballistic clouds that interact only through collisions.
The bar major axis is shown by the vertical line. 
The bar is rotating counter-clockwise.
The thick circle represents the nuclear ring.}
\label{gasorbits}
\end{figure}

\clearpage

\begin{deluxetable}{lrrrrr}
\tablecaption{Galaxy Characteristics\label{galchar}}
\tablehead{\colhead{Galaxy}&\colhead{R.A.}&\colhead{Dec.}&\colhead{Hubble}&
\colhead{Systemic} & \colhead{Linear Scale\tablenotemark{b}} \\
\omit & \colhead{J2000} & \colhead{J2000} & \colhead{Type} & \colhead{Velocity\tablenotemark{a}} &\omit}
\startdata
NGC 1300 & 3$^h$19$^m$41\fs6 &-19\arcdeg24\arcmin43\farcs1 & (R')SB(s)bc & 1552 km s\sso & 92  pc arcsec$^{-1}$\nl
NGC 1530 & 4$^h$23$^m$26\fs7 & 75\arcdeg17\arcmin43\farcs8 & SB(rs)b & 2467 km s\sso & 157 pc arcsec$^{-1}$\nl
NGC 2903 & 9$^h$32$^m$10\fs1 & 21\arcdeg30\arcmin02\farcs0 & SAB(rs)bc & 550 km s\sso &  55 pc arcsec$^{-1}$\nl
NGC 3627 &11$^h$20$^m$15\fs1 & 12\arcdeg59\arcmin21\farcs7 & SAB(s)b &726 km s\sso&  70 pc arcsec$^{-1}$\nl
NGC 4314 &12$^h$22$^m$32\fs0 & 29\arcdeg53\arcmin43\farcs3 & SB(rs)a &969 km s\sso &  81 pc arcsec$^{-1}$\nl
NGC 5135 &13$^h$25$^m$44\fs0 &-29\arcdeg50\arcmin02\farcs3 & SB(l)ab & 4112 km s\sso & 286 pc arcsec$^{-1}$\nl
NGC 5383 &13$^h$57$^m$04\fs6 & 41\arcdeg50\arcmin46\farcs0 & (R')SB(rs)b pec& 2263 km s\sso & 159 pc arcsec$^{-1}$\nl
\enddata
\tablenotetext{a}{All velocities in this table use the optical convention.}
\tablenotetext{b}{The linear scale was determined by using 
H$_0$=75 km s\sso\ Mpc\sso\ and the V$_{3k}$ from the RC3.}
\end{deluxetable}

\begin{deluxetable}{lrrlrr}
\small
\tablecaption{CO Observational Parameters\label{coobs}}
\tablehead{\colhead{Galaxy}&\colhead{Bandwidth}&
\colhead{Projected}&\colhead{T$_{sys}$}&\colhead{\# of Config. x }&\colhead{$\theta_{synth}$} \\ 
\omit&\colhead{(Total,~Imaged)}&\colhead{Baselines}&\omit&\colhead{\# of Antennas}&\omit}  
\startdata
NGC 1300 &  912,~400 \kms & 2.2 - 50 k$\lambda$ & 700-1500 K & 1x9 & 8$\farcs$7 x 3$\farcs$8 \nl
NGC 1530 &  772,~490 \kms & 2.8 - 90 k$\lambda$ & 500-1000 K & 3x6 & 5$\farcs$1 x 4$\farcs$3 \nl
NGC 2903 &  516,~400 \kms & 2.4 - 83 k$\lambda$ & 400-800 K & 3x6 & 3$\farcs$9 x 3$\farcs$5 \nl
NGC 3627 &  900,~460 \kms & 2.2 - 27 k$\lambda$ & 300-600 K & 1x9 & 7$\farcs$7 x 7$\farcs$1 \nl
NGC 4314 &  516,~420 \kms & 2.2 - 83 k$\lambda$ & 400-700 K & 3x6 & 5$\farcs$6 x 4$\farcs$2 \nl
NGC 5135 &  730,~400 \kms & 1.4 - 75 k$\lambda$ & 800-1500 K & 4x6 & 10$\farcs$2 x 3$\farcs$6 \nl
NGC 5383 &  460,~460 \kms & 2.2 - 91 k$\lambda$ & 500-1100 K & 1x9,~3x6 & 4$\farcs$6 x 4$\farcs$2 \nl
\enddata
\end{deluxetable}

\begin{deluxetable}{lrrrrr}
\tablecaption{Galaxy Position Angles\label{galposang}}
\tablehead{\colhead{Galaxy}&\colhead{Position Angle}&\colhead{Reference}}
\startdata
NGC 1300 & -93$\pm 0.5$\arcdeg& 1\nl
NGC 1530 & 8$\pm 1$\arcdeg&2\nl
NGC 2903 & 13$\pm 1$\arcdeg& 3\nl
NGC 3627 & -4$\pm 2$\arcdeg&3\nl
NGC 4314 & -25$\pm 10$\arcdeg&4\nl
NGC 5135 & -56$\pm 2$\arcdeg&3\nl
NGC 5383 & 85\arcdeg&5\nl
\tablerefs{(1) Lindblad et al. 1997\nl
(2) Regan et al. 1996\nl
(3) This Paper\nl
(4) Benedict et al. 1996\nl
(5) Duval \& Athanassoula 1983}
\enddata
\end{deluxetable}
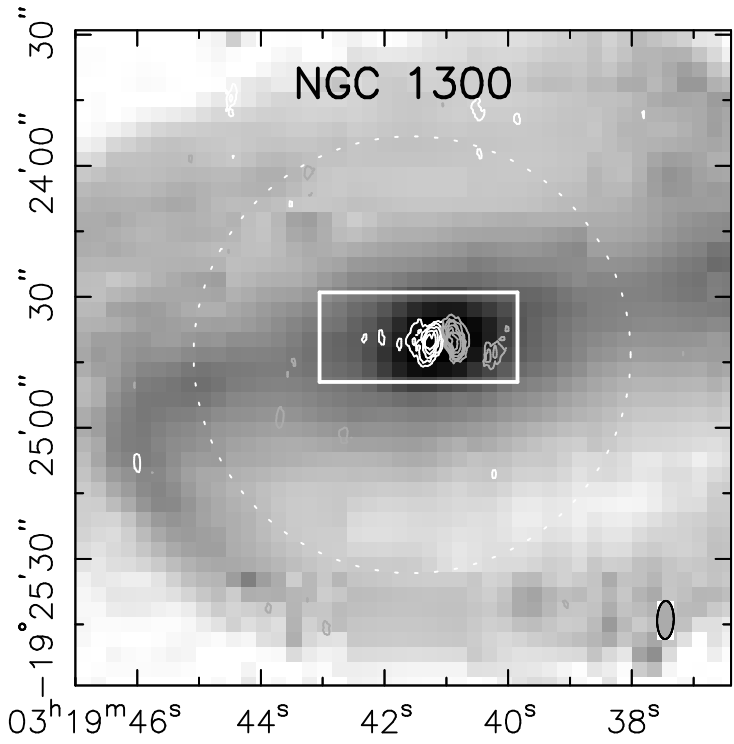
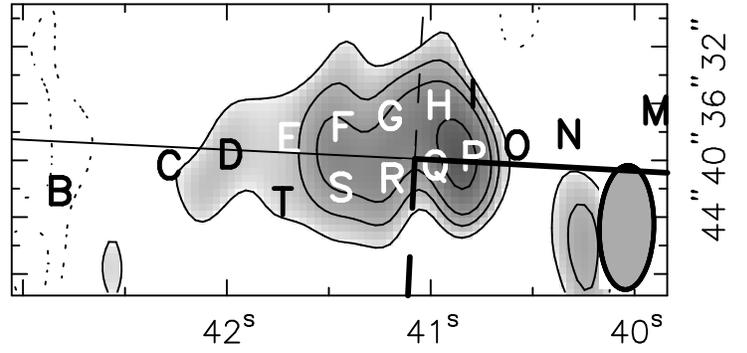
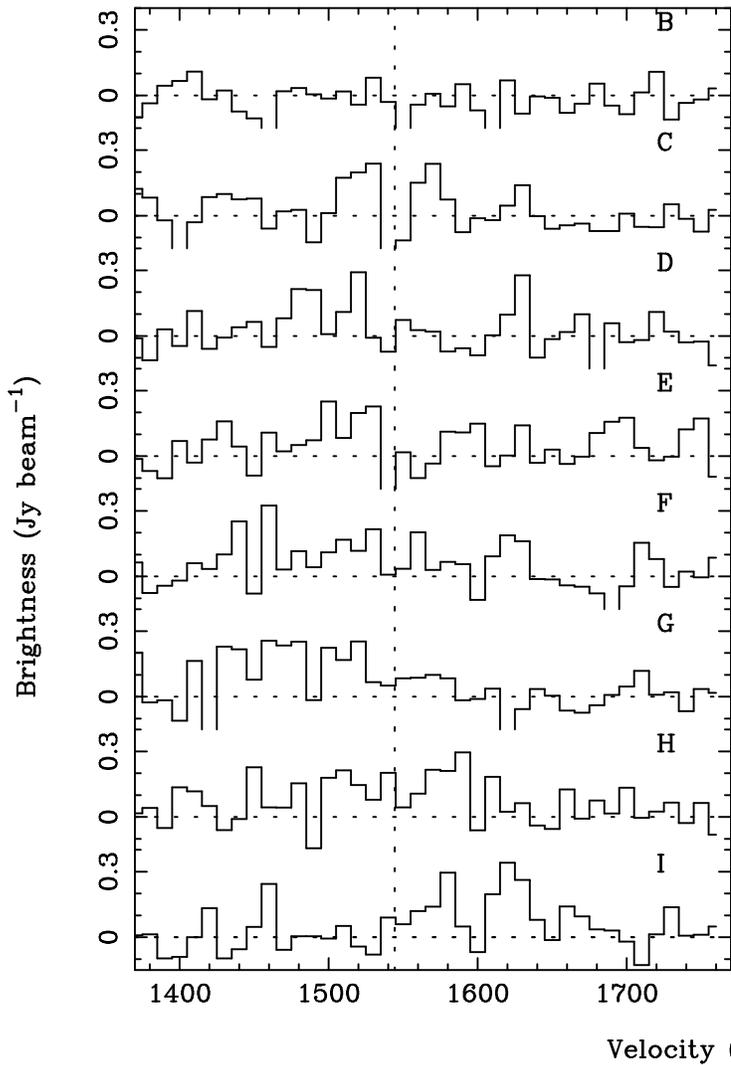
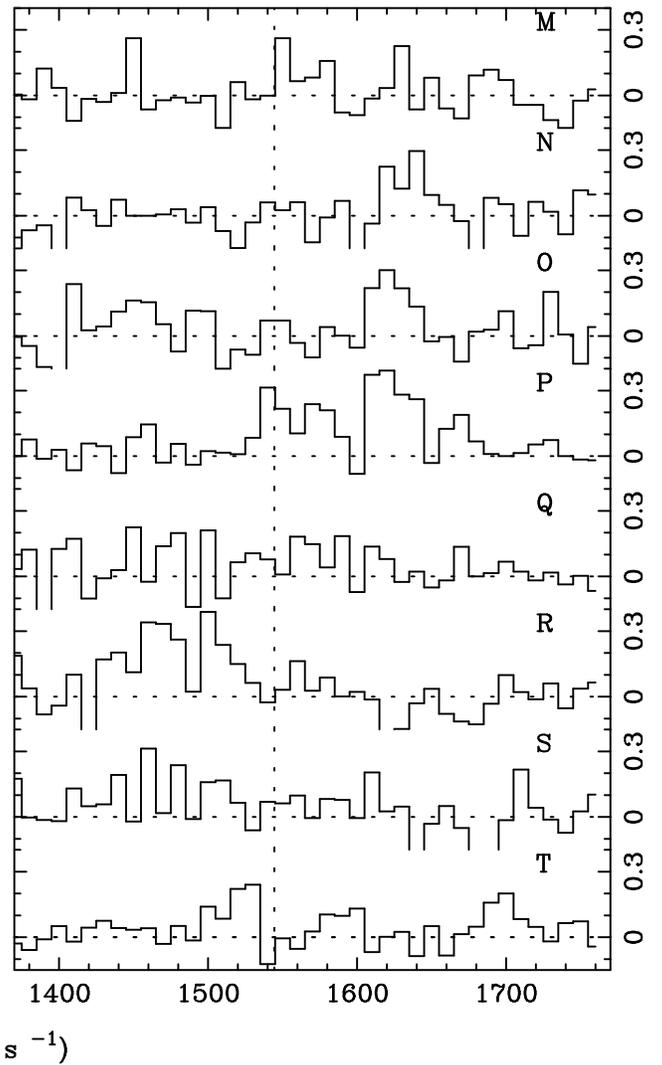
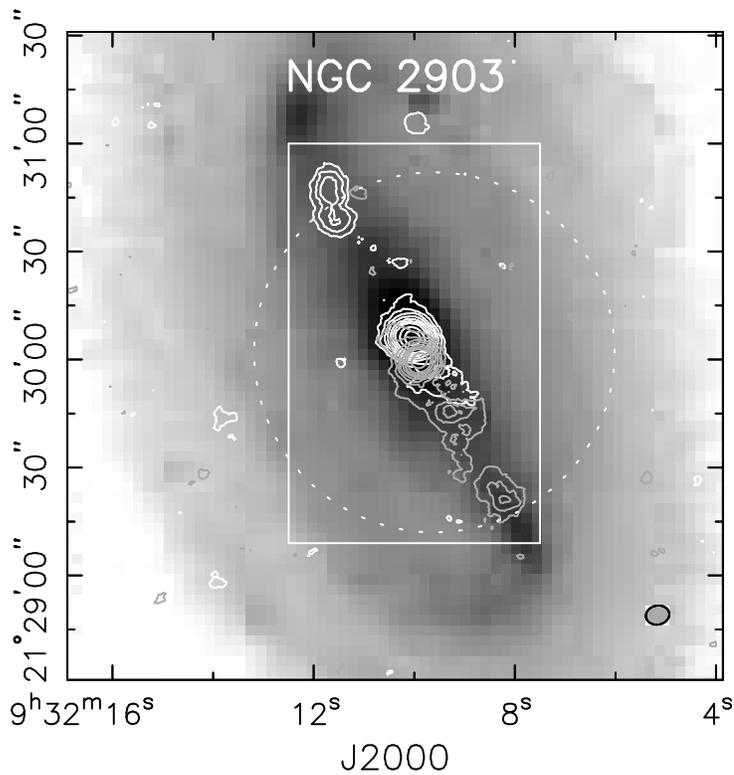
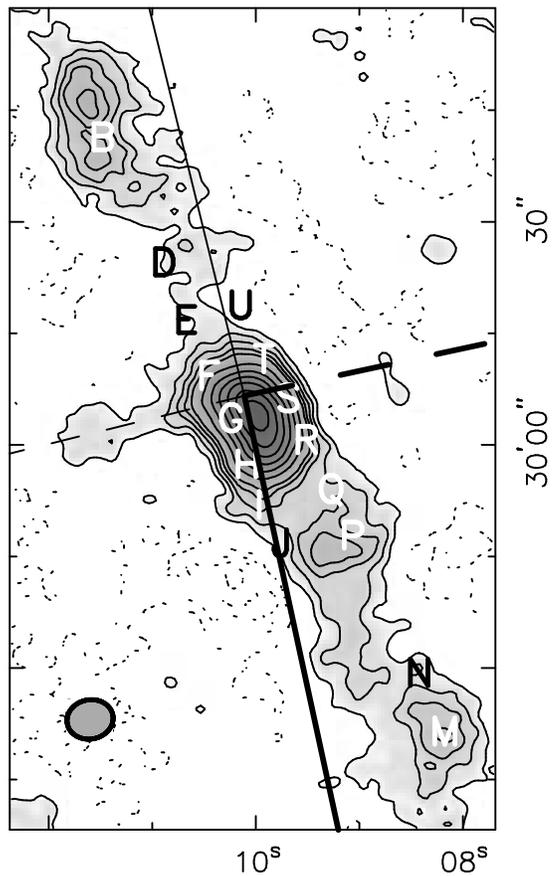
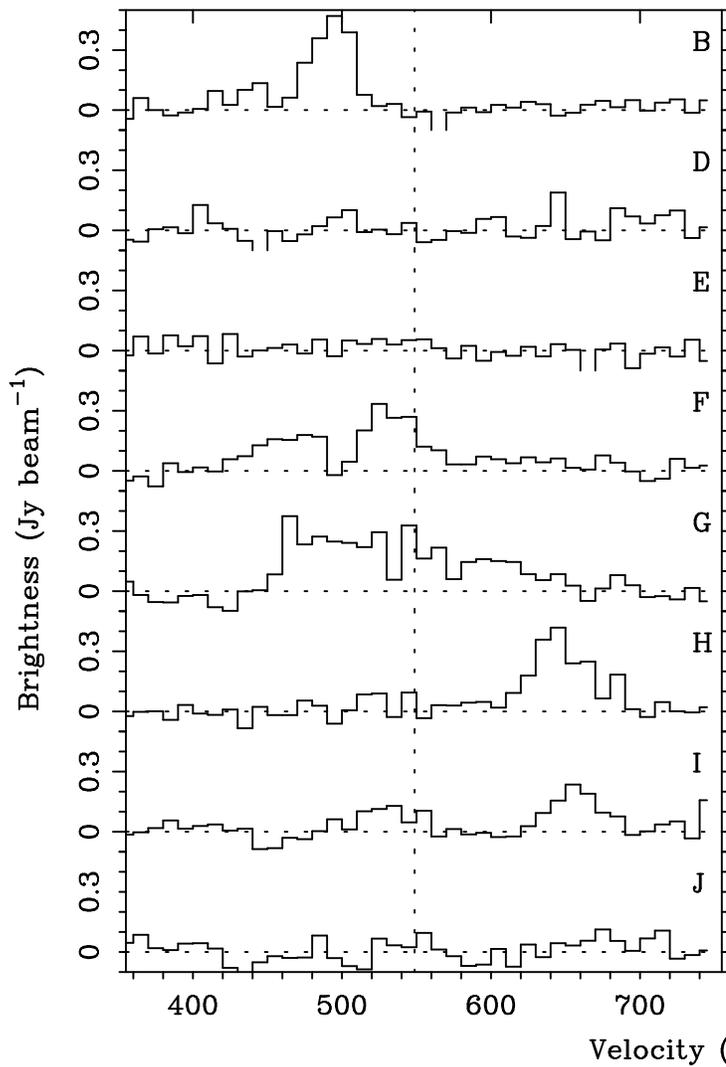
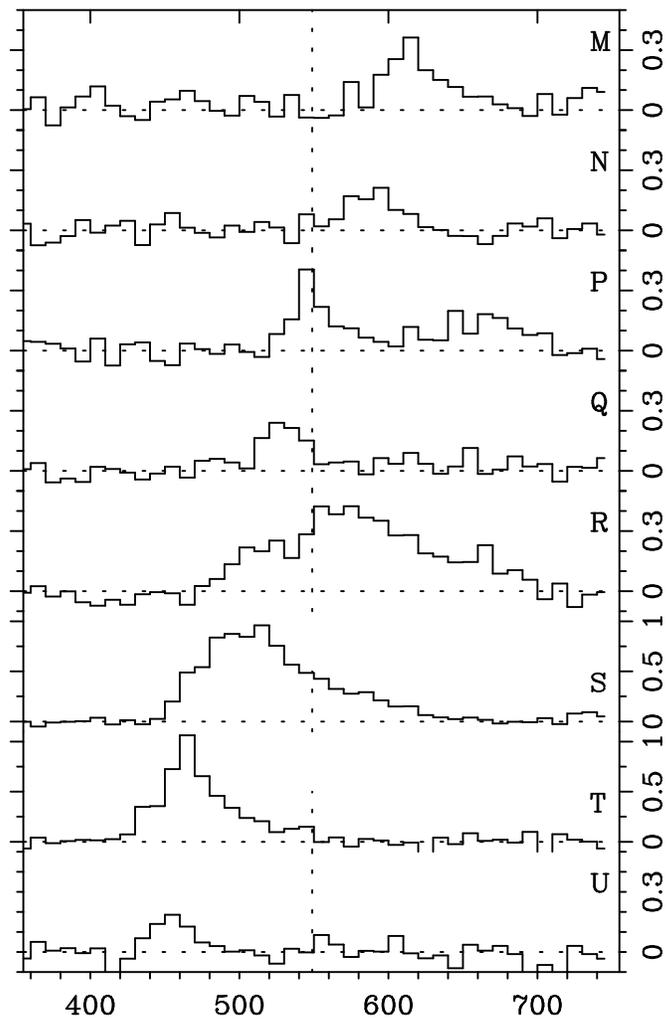
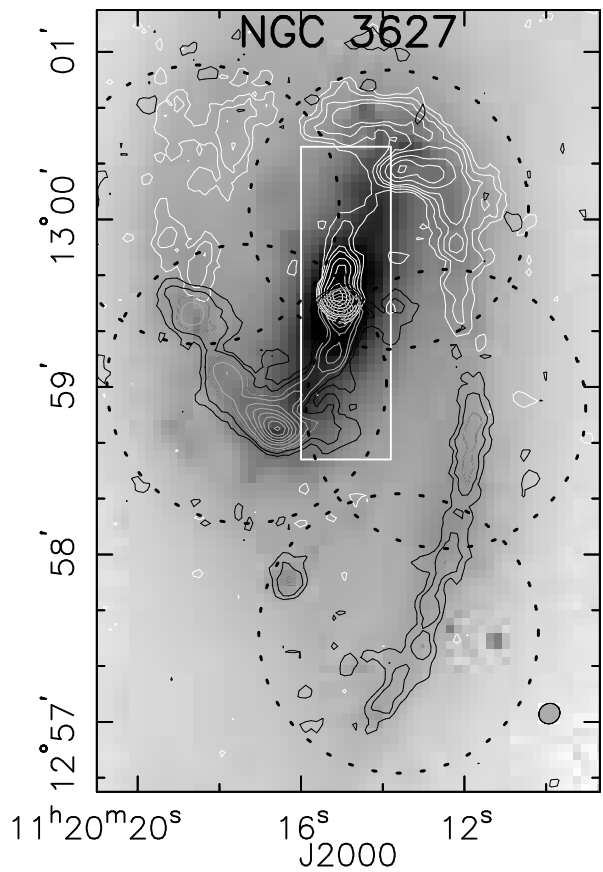
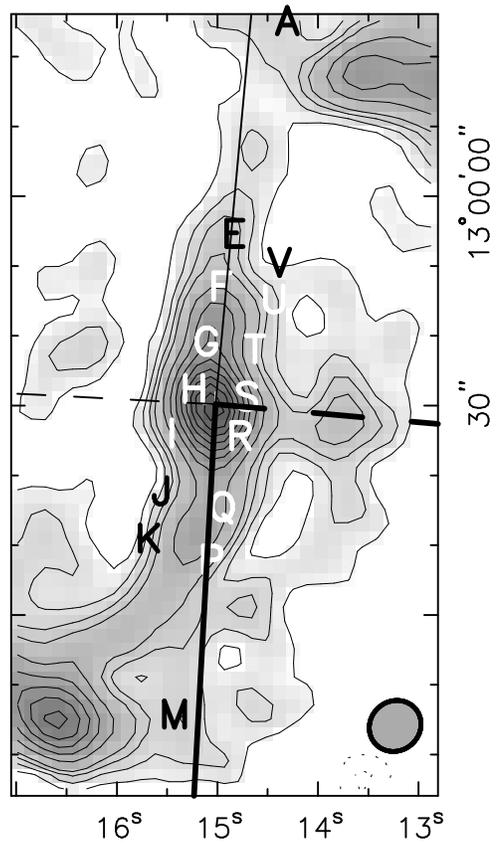
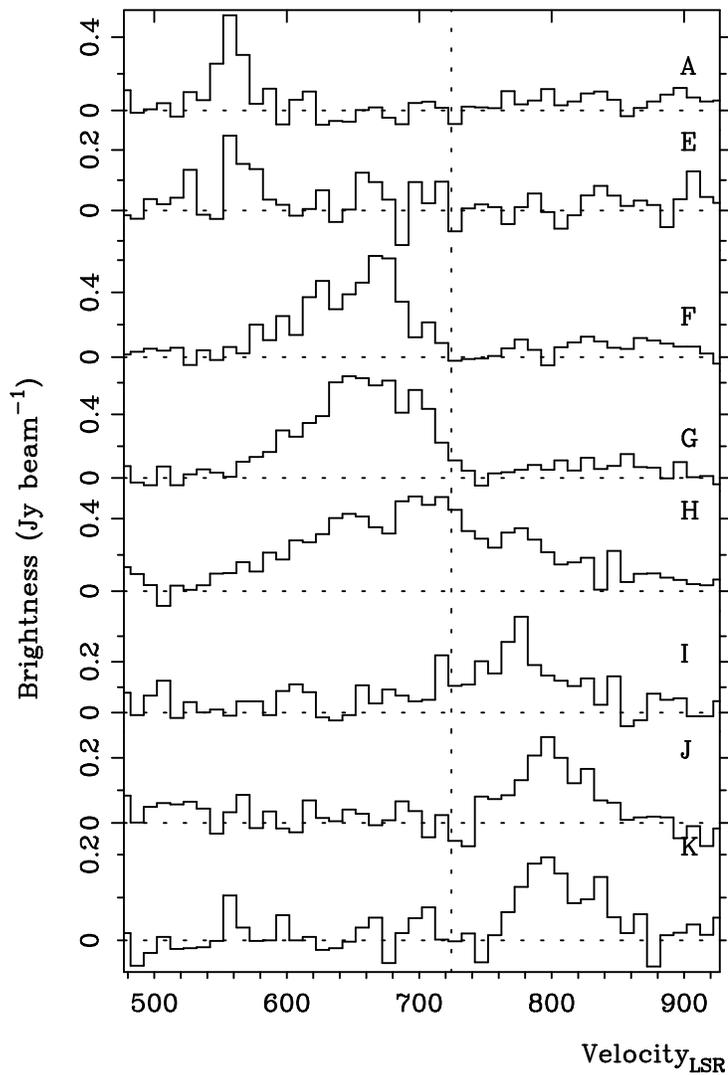
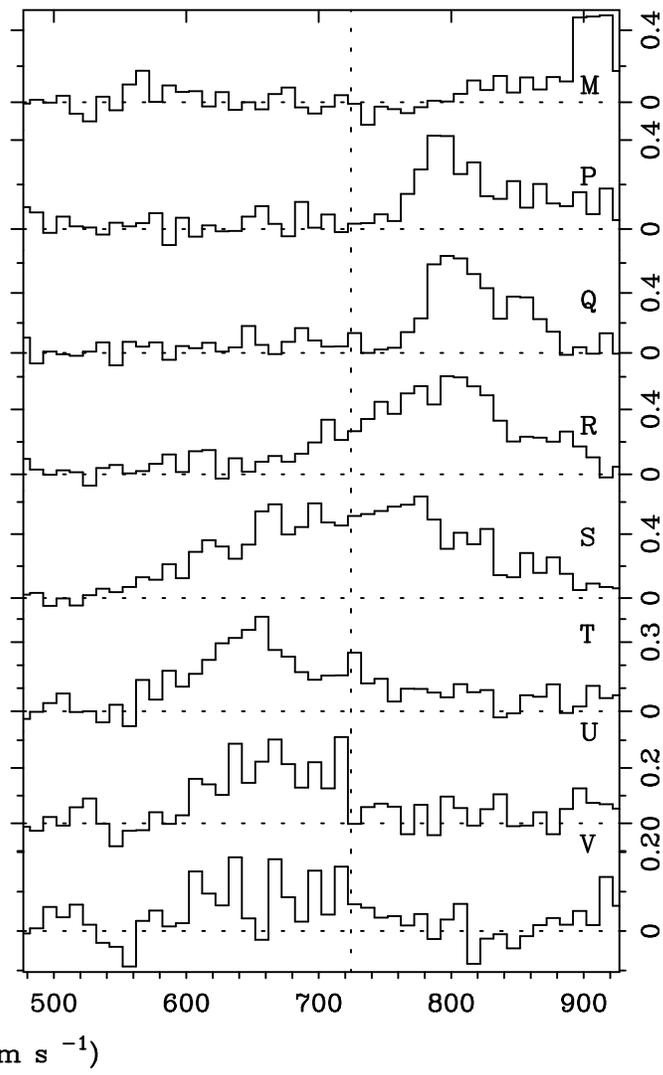
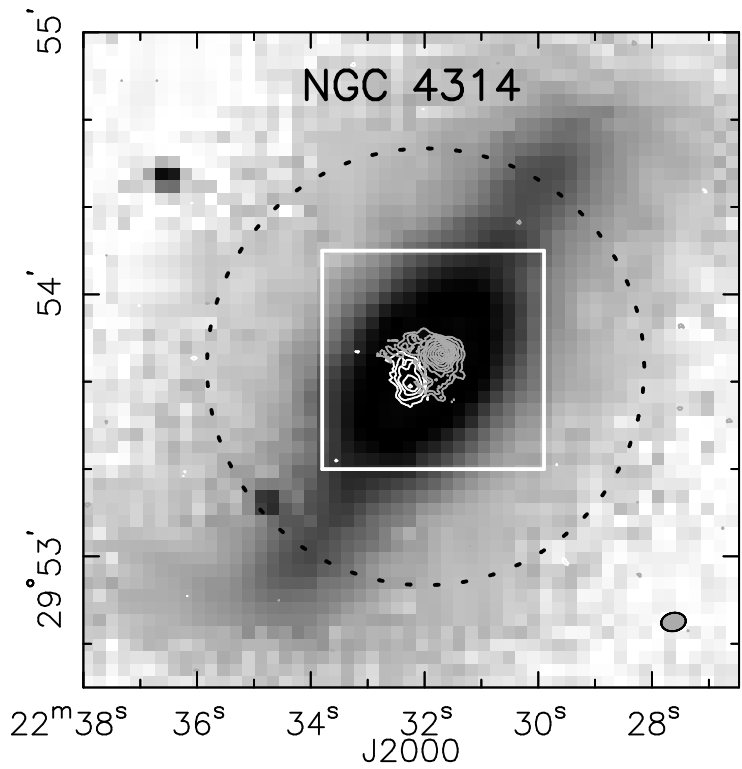
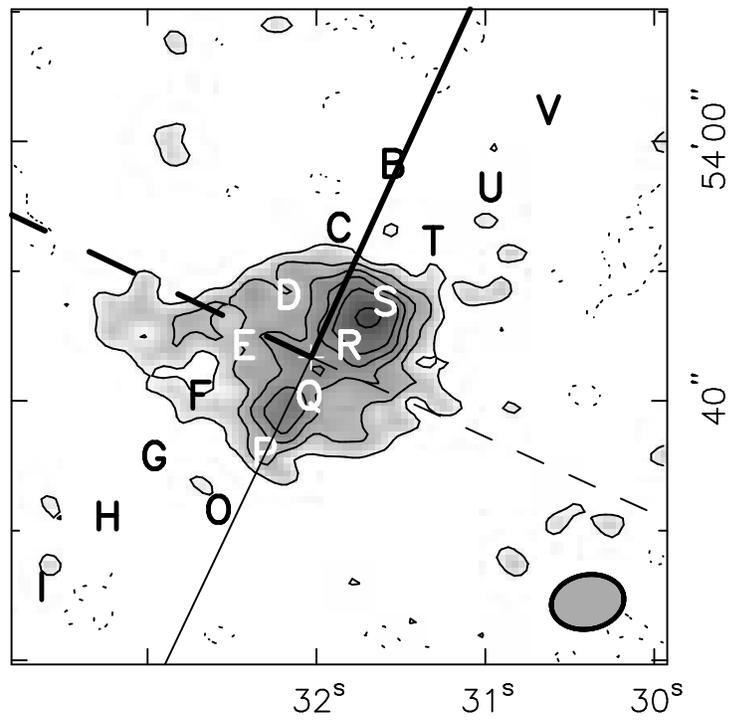
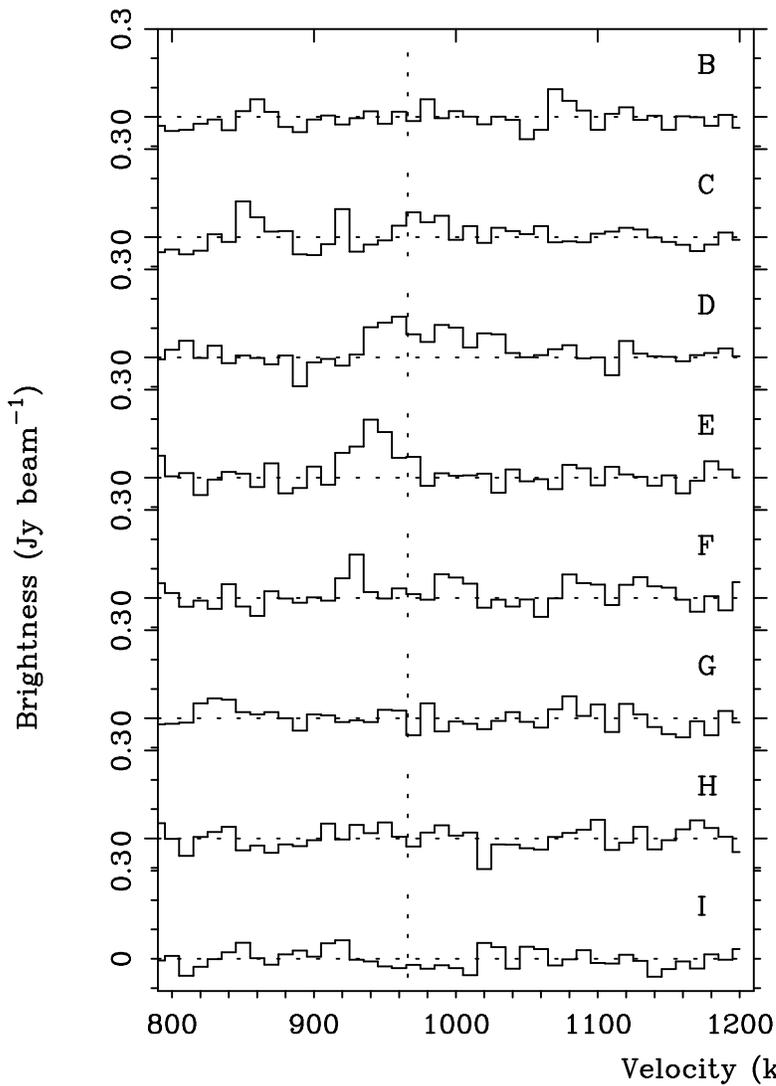
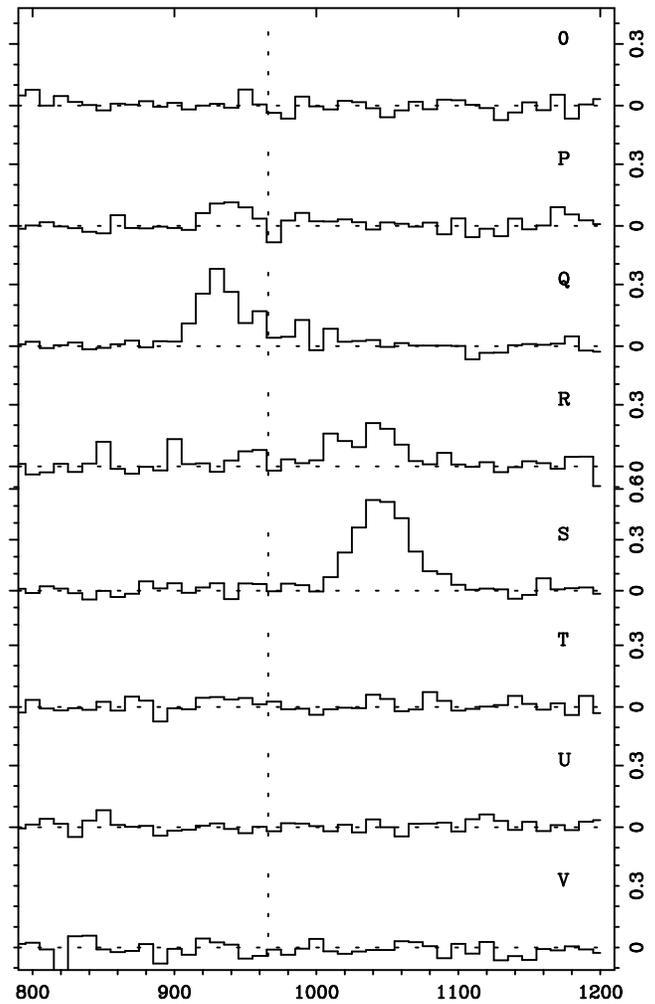
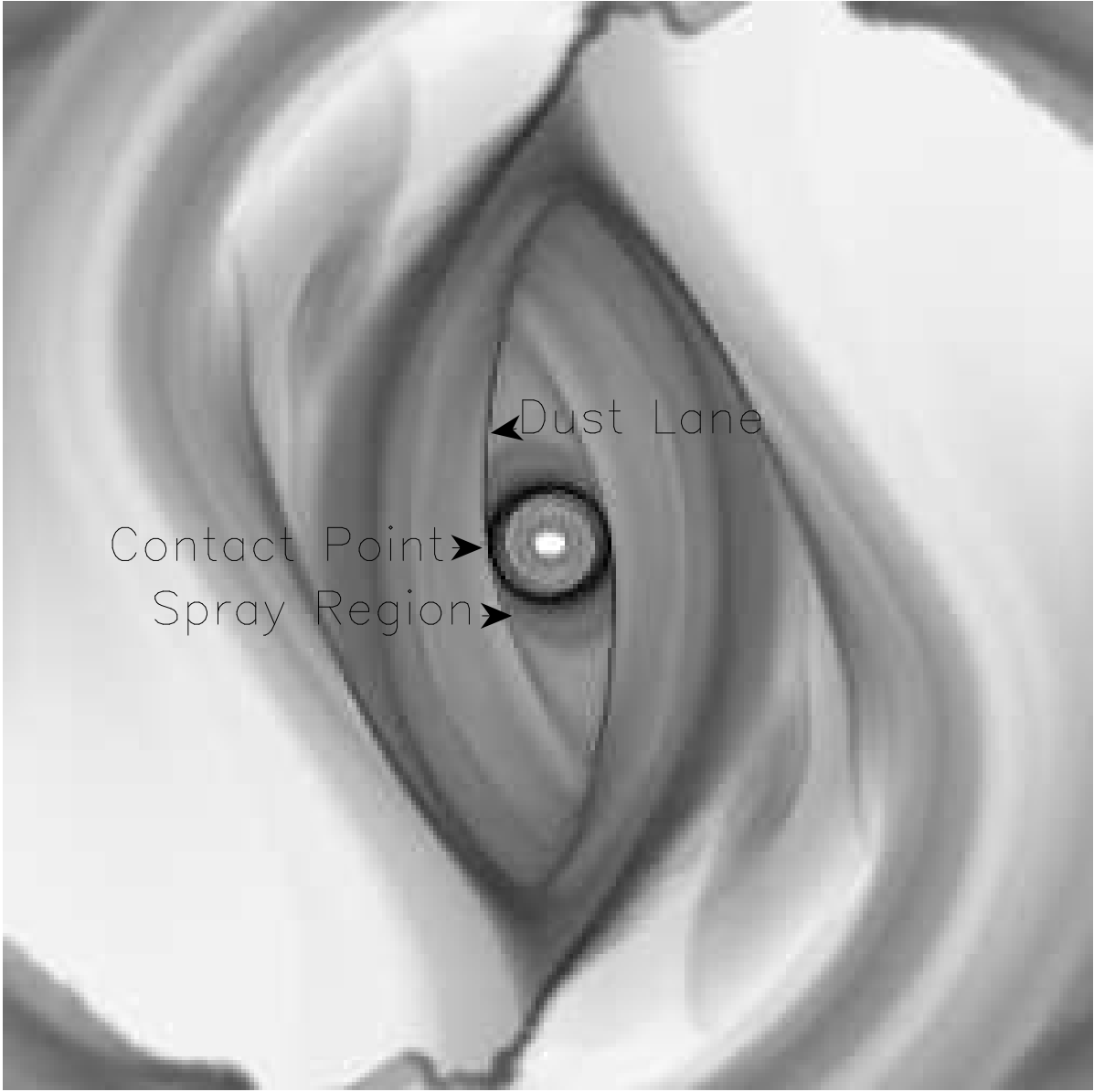
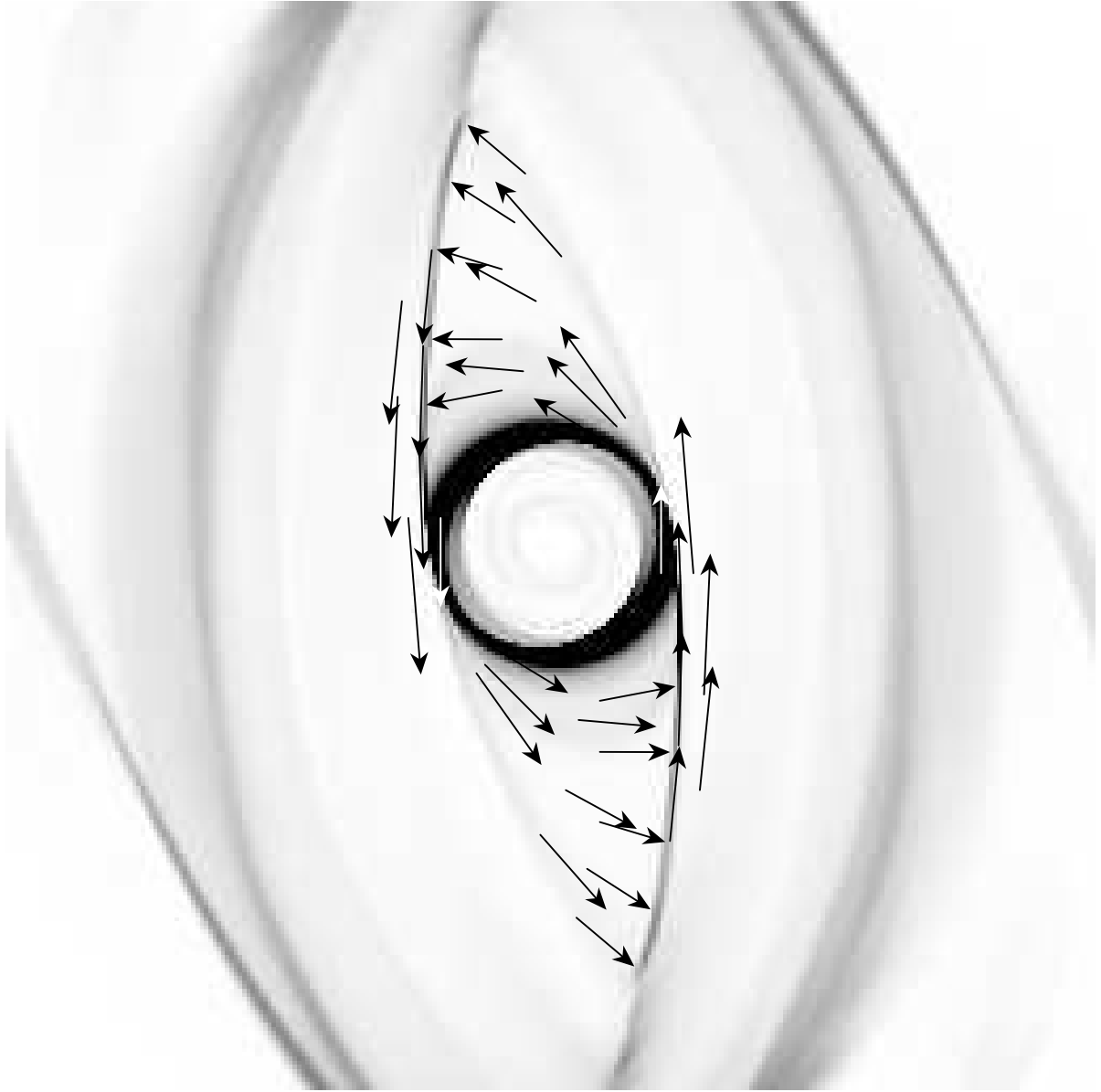
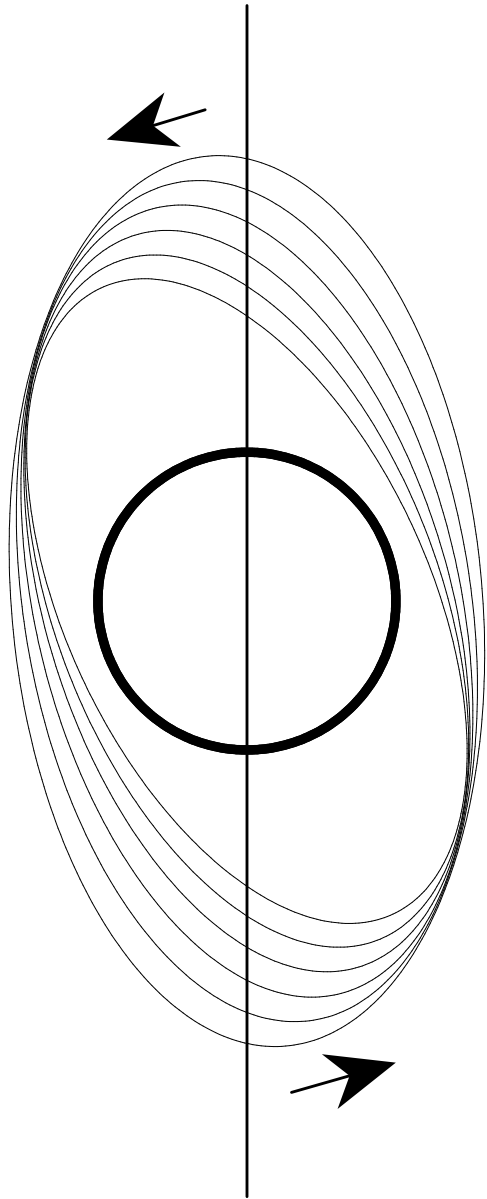
\end{document}